\documentclass[a4paper]{article}
\usepackage{RR}
\RRNo{0338}
\usepackage{color}
\usepackage{hyperref}
\usepackage{listings}
\usepackage{longtable}
\usepackage{lscape}
\hypersetup{colorlinks=true, linkcolor=blue, filecolor=blue, pagecolor=blue, urlcolor=blue}
\DeclareFixedFont{\mytabfont}{\encodingdefault}{\familydefault}{\seriesdefault}{\shapedefault}{8pt}
\newcommand{\tabfont}{\mytabfont}
\newcommand{\paf}{\vspace{0.5em}\par}
\RRdate{June 2007}
\RRauthor{Yvan Royon\and St\'ephane Fr\'enot}
\authorhead{Y. Royon \& S. Fr\'enot}
\titlehead{A Survey of Unix Init Schemes}
\RRtitle{Etat de l'art des processus init}
\RRetitle{A Survey of Unix Init Schemes}
\RRresume{

Dans la plupart des syst\`emes d'exploitation modernes, init (comme
``initialisation'') est l'application lanc\'ee par le noyau au d\'emarrage.
Il s'agit d'un d\'emon qui a le PID 1. init a pour t\^aches de lancer
les autres processus et de nettoyer les processus zombies. Il est aussi
responsable du d\'emarrage et red\'emarrage du syst\`eme.

Ce document d\'ecrit les implantations existantes du processus
init et des scripts init pour variantes d'Unix.
Ces implantations vont des solutions BSD et SystemV, pionni\`eres et
toujours utilis\'ees, \`a celles plus r\'ecentes et prometteuses venant de
Ubuntu, Apple, Sun, ainsi que de d\'eveloppeurs ind\'ependants.
Notre but ici est d'analyser les objectifs de ces implantations, et
de comparer leurs fonctionnalit\'es.

}
\RRabstract{

In most modern operating systems, init (as in ``initialization'') is the
program launched by the kernel at boot time. It runs as a daemon and
typically has PID 1. Init is responsible for spawning all other
processes and scavenging zombies. It is also responsible for reboot and
shutdown operations.

This document describes existing solutions that implement the init process
and/or init scripts in Unix-like systems.
These solutions range from the legacy and still-in-use BSD and SystemV
schemes, to recent and promising schemes from Ubuntu, Apple, Sun and
independent developers.
Our goal is to highlight their focus and compare their sets of features.

}
\RRmotcle{processus init}
\RRkeyword{init process, init scheme}
\RRprojet{Ares}
\RRtheme{\THCom}
\URRhoneAlpes
\begin{document}
\makeRT
\newpage

%%%%%%%%%%%%%%%%%%%%%%%%%%%%%%%%%%%%%%%%%%%%%%%%%%%%%%%%%%%%%%%%
%%%%%%%%%%%%%%%%%%%%%%%%%%%%%%%%%%%%%%%%%%%%%%%%%%%%%%%%%%%%%%%%

\section*{The Boot Process}
What happens between the moment a computer is switched on and the moment a
user can log in?

\paf Let us begin by a quick and simplified overview of the boot process
in most modern operating systems.

\begin{enumerate}

\item The computer is switched on or rebooted;

\item The \textbf{BIOS}, stored on the motherboard, is executed from a known
location (flash memory);

\item The BIOS determines the boot device (local disc, removable media,
PXE from the network);

\item The boot device has a special, known sector called the Master Boot
Record, which contains the \textbf{stage 1 bootloader}. The stage 1
bootloader is loaded into RAM, and points to the stage 2 bootloader. The
latter can be several sectors long, and is stored on the boot device in a
location unknown to the BIOS and subject to change;

\item The \textbf{stage 2 bootloader} (e.g., LILO or GRUB for Linux)
uncompresses and launches the kernel;

\item The \textbf{kernel} runs a single application. For Linux, the
default is \textsf{/sbin/init}, and can be modified in LILO or GRUB by
passing the \textsf{init=/path/to/executable} parameter to the kernel.

\end{enumerate}

Usually, \textsf{init} launches a few \textsf{getty} processes for
virtual terminals, a few daemons (cron, dhcp, acpi), the X server, etc.

\vspace{1em}

On Unix-like systems, \textsf{init} (as in ``initialization'') is the
program launched by the kernel at boot time. It runs as a daemon and
typically has PID 1. \textsf{init} is responsible for spawning all other
processes and scavenging zombies. It is also responsible for reboot and
shutdown operations.

\section*{A Bit of History}

The first widespread \textsf{init} is the simple solution from BSD. Its
main competitor is the more featureful but more complex System V
\textsf{init}.

The BSD \textsf{init} process simply launches a list of processes.  It
basically reads a configuration file (\textsf{/etc/rc}) where all programs
to launch at startup are statically listed. If the administrator wants to
modify this list, she must edit the file manually, which is clearly
error-prone and has potentially disastrous effects. This is a very simple
and lightweight scheme, but it is completely static.

System V, on the other hand, chose to be more flexible. It introduces the
notion of \emph{runlevel}, which can be viewed as multiple and specialized
BSD \textsf{rc} files. A runlevel has an identifier and a purpose.
Runlevel 1 boots into single mode, runlevel 2 boots into multi-user mode,
etc. Only a fixed number of eight runlevels are available.
What each runlevel does is specified in the \textsf{/etc/inittab} file.

A runlevel is a \emph{software configuration} of the system which allows
only a selected group of applications to exist. When \textsf{init} is
requested to change the runlevel, it kills all applications that are not
listed in the new runlevel, and starts all unstarted applications in the
new runlevel.

The second great improvement in SysVinit is the notion of rc script. Each
application contained in a runlevel comes with a wrapper script; it is an
executable that accepts \textsf{start} and \textsf{stop} as parameters.
This allows to launch and kill each application using the same syntax,
regardless of how they are implemented and how they handle parameters
internally.

Most often, the combination of rc scripts with runlevels makes use of
symbolic links.  A runlevel is a directory that contains symlinks named
with a numbered prefix. When entering a runlevel, symlinks are accessed
alphabetically, in the static order defined by their number. This allows
to reuse the same rc script in multiple runlevels.

\section*{Evolution of Init Schemes}

Today, most GNU/Linux distributions use a derivative of System V's
\textsf{init}. Even NetBSD went from the BSD \textsf{init} to a
SystemV-like \textsf{init}. However, there is room for lots of
improvements, and lots of proposals exist on several topics.\\

The first topic is the order in which rc scripts are started. In 2002, the
\textsf{need(8)} scheme proposed to drop statically, manually ordered
lists in favour of \textbf{dependencies}. Each rc script lists the
services it needs (e.g., the DHCP daemon needs a network connection), and
\textsf{init} computes which rc scripts must be started, in what order.
All init schemes except BSD, System V and three minor others have a way to
express dependencies.

The second topic is the enhancement of \textbf{configuration management}.
For instance, Gentoo rc, initNG and runit identify runlevels using a name
instead of a number. This breaks the limitation on the number of
runlevels, and allows to dedicate a runlevel to a context or a particular
usage (e.g., "AC mode" and "battery mode" on a laptop).  Similarly,
upstart, cinit and Sun SMF introduce a notion of profile (or "mode" for
einit).

Another topic is \textbf{life cycle management}. There are two main
solutions when monitoring the life cycle of applications. The first one is
when using rc scripts: \textsf{init} (or a separate \textsf{rc}
application) use the \textsf{/var} directory to store which daemon has
been started, with which PID. If no process with such PID is running, then
the daemon has been killed. This is a reactive life cycle management,
since there is a delay between the death of the daemon and \textsf{init}
or \textsf{rc} noticing it. This scheme is used in System V, Gentoo rc,
simpleinit, initNG, minit, cinit, monit and the LSB specifications. We
must also note that rc scripts may differ for each distribution and init
scheme: they handle \textsf{/var} differently.
The other way to monitor the life cycle of daemons is not to use rc
scripts. A "universal father" process forks and execs daemon processes
directly. The pro is that it is  a proactive life cycle management: POSIX
signals are used whenever a daemon dies. The con is that the daemon must
handle itself portability (the environment variables and the parameters it
uses), instead of delegating it to an external script. Sun SMF, launchd,
upstart and einit use such a "universal father". runit and svscan use a
different "father" process for each daemon.

Init schemes allow a number of operations to \textbf{instrument} the life
cycle of applications. Solutions that use father processes allow mostly
what POSIX signals handle: start, stop, and a hook to execute when a
daemon dies unexpectedly. This hook can be used to log errors or to
automatically restart ("respawn") the faulty daemon. Solutions that use rc
scripts may allow a restart operation on a script (Gentoo rc, monit,
NetBSD rc.d, LSB specifications), or respawn (initNG, minit, jinit,
cinit), in addition to start and stop (System V and its derivatives).

Yet another topic for init enhancement is the \textbf{tools} that ease
development and usage. System V uses \textsf{telinit} to pass commands to
\textsf{init}. Debian introduced \textsf{start-stop-daemon}, that handles
PID files and such. Gentoo uses \textsf{rc-update} to create and modify
runlevels. In Sun SMF, \textsf{svccfg} switches profiles. Most init
schemes implement their own specific tools.

Finally, the great number of existing init schemes is explained by their
different focus. Some focus on startup speed (initNG parallelizes startup
scripts). Some focus on implementation size (minit, cinit), in order to be
usable on resource-constrained devices. Others focus on one of the topics
listed above (dependencies, runlevels, life cycle management and
instrumentation, tools). But the most interesting schemes are the fairly
new ones backed up by Apple, Sun and Ubuntu.\\

Apple's launchd aims to replace all programs that start applications:
\textsf{xinetd}, \textsf{crond}, \textsf{anacron} and \textsf{init}.
Daemons can be started, respawned, etc, but also scheduled.

Sun's SMF provides extensive failure detection and diagnostics. It allows
to run services in degraded or maintenance mode. The goal is to ease the
maintenance of the system (presumably a big server) and to enhance uptime.

Ubuntu's upstart aims to replace the same tools as launchd, but also to
unify devices management and power management. It transforms anything into
events and triggers: a new device appearing is an event, the screen going
into sleep mode is an event, etc. Actions (e.g., mount a filesystem) can
be triggered when specific events are thrown. Daemons and dependencies are
handled the same way: the DHCP daemon cannot start until a "network is
ready" event has been thrown.

% Conclusion?

\section*{Technical Summaries}

The remainder of this document are technical summaries of all init-related
schemes that the authors know of. Each summary gives a short description
on the init scheme, its goals and/or its origin. It then lists features in
terms of configuration, life cycle, dependency and service management.

%%%%%%%%%%%%%%%%%%%%%%%%%%%%%%%%%%%%%%%%%%%%%%%%%%%%%%%%%%%%%%%%

\clearpage
\setcounter{tocdepth}{1}
\tableofcontents

%%%%%%%%%%%%%%%%%%%%%%%%%%%%%%%%%%%%%%%%%%%%%%%%%%%%%%%%%%%%%%%%

\clearpage\section{BSD init}
\label{sec:bsd}
\subsection{Home Page}
Documentation for recent versions of the BSD-style \textsf{init}:
\\
\url{http://www.freebsd.org/doc/en\_US.ISO8859-1/books/handbook/boot-init.html}
\\
The source code for FreeBSD and OpenBSD can be found at :
\\
\url{http://www.freebsd.org/cgi/cvsweb.cgi/src/sbin/init/}
\\
\url{http://www.openbsd.org/cgi-bin/cvsweb/src/sbin/init/}

\subsection{About}

\textsf{/sbin/init} runs a master boot script (\textsf{/etc/rc}) which
orchestrates the whole boot procedure. There may be a few additional
\textsf{rc.*} scripts. \textsf{init} launches them before going in
multi-user mode. Because the boot process is sequential and has very few
scripts, it is simple and fast.

However, if a new package needs to register itself into the boot
process, it needs to edit one of the existing script files. This is
very dangerous: if the package installer makes a mistake, the whole
startup process will fail.

\subsection{Configuration Management}

BSD init only handles one or a few boot scripts when switching from single-user
mode to multi-user mode, and with \textsf{halt} / \textsf{reboot} /
\textsf{shutdown} operations. There are no profiles or runlevels.

\subsection{Life Cycle Management}
None.

\subsection{Dependency Resolution}
No dependencies; only an ordered, linear startup defined in
\textsf{/etc/rc}.

\subsection{Service Management Model}
None.

\subsection{Service Management Tools}
None. Unix tools such as \textsf{ps} or \textsf{signal} are not part of
\textsf{init}.
\\
Communication with the outside world is achieved through signals.

\clearpage\section{System V init }
\label{sec:sysv}
\subsection{Home Page}
\url{http://freshmeat.net/projects/sysvinit/}

\subsection{About}
From the Freshmeat page:

Init's primary role is to create processes from a
script stored in the \textsf{/etc/inittab} file. This package also
contains well known and used utilities like \textsf{reboot},
\textsf{shutdown}, \textsf{killall}, \textsf{poweroff},
\textsf{telinit}, \textsf{sulogin}, \textsf{wall}, etc.

\subsection{Configuration Management}
SysVinit introduced the notion of runlevel. From the \textsf{man} page:

A runlevel is a software configuration of the system which allows only a
selected group of processes to exist. The processes spawned by
\textsf{init} for each of these runlevels are defined in the
\textsf{/etc/inittab} file. Init can be in one of eight runlevels:
0-6 and S. The runlevel is changed by having a privileged user run
\textsf{telinit}, which sends appropriate signals to \textsf{init},
telling it which runlevel to change to.

Runlevels 0, 1, and 6 are reserved. Runlevel 0 is used to halt the
system, runlevel 6 is used to reboot, and runlevel 1 is used to get the
system down into single user mode. Runlevel S is used before entering
runlevel 1 (single-user mode). Runlevels 2-5 are multi-user
modes, and are different for each Unix / Linux distribution.

%When \textsf{init} is requested to change the runlevel, it kills all
%processes that are undefined in the new runlevel (via SIGTERM first,
%then SIGKILL if still alive).
%
%Some variants (e.g. the Linux version found in RedHat) use directories
%to define runlevels. From nico.schotteli.us:

When entering a new runlevel (at system boot the runlevel is undefined,
sometimes described as ``N''), SysVinit executes a master script
(\textsf{/etc/init.d/rc} for instance). This script executes all
stop-scripts from the old runlevel, then all start-scripts from the new
runlevel. Those scripts are found either in \textsf{/etc/init.d},
\textsf{/etc/rc.d} or \textsf{/sbin/init.d}.  Each runlevel has a
directory: \textsf{/etc/rc2.d}, \textsf{/etc/rc6.d}, etc. Those
directories contain links to the real scripts. SysVinit detects whether to
stop or to start a script by its name: start-scripts begin with a ``S'',
stop-scripts with a ``K'' (as in ``kill''). When starting the
start-scripts, they are called with ``start'' as the first and only
parameter. When starting stop-scripts, they are called with ``stop'' as
first parameter. This is how a configuration for runlevel 2 could look
like:

\vspace{0.5em}
\lstset{language=sh, basicstyle=\footnotesize}
\begin{lstlisting}[frame=single]
scice% ls -l /etc/rc2.d
total 0
lrwxr-xr-x  1 root root 18 Feb 20  2004 S10sysklogd -> ../init.d/sysklogd
lrwxr-xr-x  1 root root 15 Feb 20  2004 S11klogd -> ../init.d/klogd
lrwxr-xr-x  1 root root 13 Feb 20  2004 S14ppp -> ../init.d/ppp
lrwxrwxrwx  1 root root 15 Feb 24  2004 S15bind9 -> ../init.d/bind9
lrwxrwxrwx  1 root root 18 Feb 20  2004 S18quotarpc -> ../init.d/quotarpc
[...]

scice% ls -l /etc/rc6.d
lrwxrwxrwx  1 root root 15 Aug 31 16:22 /etc/rc6.d/K19aumix -> ../init.d/aumix
lrwxrwxrwx  1 root root 15 Apr 19 09:55 /etc/rc6.d/K19samba -> ../init.d/samba
lrwxrwxrwx  1 root root 13 Feb 25  2004 /etc/rc6.d/K20gpm -> ../init.d/gpm
[...]
\end{lstlisting}

(Source: \url{nico.schotteli.us})

The number behind the S or the K is the priority. The lower the
priority, the earlier the script is started.

Each service has its own init-script, which sometimes also accepts
``restart'' or ``reload''. This way, runlevels are modular.

\subsection{Life Cycle Management}

The \textsf{/var/log/wtmp} binary file records all user logins and
logouts. It is maintained by \textsf{init}, \textsf{login} and
\textsf{getty}.

The \textsf{/var/run/utmp} binary file holds information on users and
processes currently using the system. Not all applications use it, so it
may be incomplete.

Also, service scripts in \textsf{/etc/init.d} can be called during
runtime from the command line (not only at startup).

SysVinit can dump its state and exec a later version of itself, so no
state is lost between upgrades (source: jinit's page).

\subsection{Dependency Resolution}
None. Services are started in alphabetical order, hence the priority
numbers in the script names. We just hope the administrator made no
mistakes\ldots

\subsection{Service Management Model}
Wrapper scripts are called with a \textsf{start} or \textsf{stop} parameter.

\subsection{Service Management Tools}
\textsf{telinit} switches the current runlevel.

\clearpage\section{The need(8) Concept }
\label{sec:need}
\subsection{Home Page}
\url{http://www.atnf.csiro.au/people/rgooch/linux/boot-scripts/}

\subsection{About}
Richard Gooch wrote this whitepaper in 2002. He argues that BSD init is
not scalable (modifications are risky), and that SysVinit is too
complex and ugly.

Gooch proposes to remove all master scripts and numbering schemes.
\textsf{init} runs all scripts in \textsf{/etc/init.d}, in random (?)
order. The scripts themselves express their dependencies with the
\textsf{need} program. This ensures that a correct startup order is
automatically observed. \textsf{need} also ensures that all services
are run only once.

The concept of virtual service names is also introduced. For instance,
\textsf{sendmail} and \textsf{qmail} both provide the \textsf{mta}
service. Which implementation is used is not important, so other
services simply \textsf{need}
``\textsf{mta}''.
This is done with the \textsf{provide} program.

These ideas are implemented in \textsf{simpleinit}.

\subsection{Configuration Management}
A runlevel is represented by a script, e.g.
\textsf{/sbin/init.d/runlevel.3}, which will \textsf{need} an unordered
set of services, as well as e.g. \textsf{runlevel.2}. This makes it
easy to increment and decrement the runlevel.

However, this does not support runlevels that bypass these precedence
relations.

\subsection{Life Cycle Management}
Same as SysVinit.

\subsection{Dependency Resolution}
Simplistic dependency expression: \textsf{need serviceXX}.

It is unclear how we ensure that \textsf{provide}, called in service
wrapper scripts, is run before dependency resolution.

\subsection{Service Management Model}
Same as SysVinit.

\subsection{Service Management Tools}
None.

\clearpage\section{Gentoo rc }
\label{sec:gentoo}
\subsection{Home Page}
\url{http://www.gentoo.org/doc/en/handbook/handbook-x86.xml?part=2&chap=4}

\subsection{About}
Gentoo uses SysVinit with modified scripts and tools. Scripts are not
handled directly by \textsf{init}, but by Gentoo's
\textsf{/sbin/rc}.

Runlevels have names instead of numbers; this removes the limit on their
number, and allows specialized runlevels such as ``unplugged'' for a
laptop. Additional tools allow to create, delete, modify runlevels from
the command line.

A global option allows to start scripts in parallel, for a performance
gain (more visible on multi-processor or multi-core machines).

Like in SysVinit, after \textsf{init} is invoked as the last step of the
kernel boot sequence, it reads \textsf{/etc/inittab}. First, the
\textsf{sysinit} entry is run:

\lstset{language=sh, basicstyle=\footnotesize}
\begin{lstlisting}[frame=single]
si::sysinit:/sbin/rc sysinit
\end{lstlisting}

\textsf{rc} is called and mounts all filesystems. Second, the
\textsf{bootwait} entry is run:

\lstset{language=sh, basicstyle=\footnotesize}
\begin{lstlisting}[frame=single]
rc::bootwait:/sbin/rc boot
\end{lstlisting}

Here, \textsf{rc} switches the runlevel to \textsf{boot}. All links in the
\textsf{/etc/runlevels/boot} directory correspond to services that
always must be run at startup: checkroot, checkfs, clock, etc. Lastly,
the \textsf{initdefault} entry is run:

\lstset{language=sh, basicstyle=\footnotesize}
\begin{lstlisting}[frame=single]
id:3:initdefault:
l3:3:wait:/sbin/rc default
\end{lstlisting}

Here, \textsf{initdefault} indicates that the default runlevel is 3 (as
in the old SysVinit scheme), and runlevel 3 says that it will wait
until \textsf{/sbin/rc} has finished switching to runlevel
\textsf{default}.

Once \textsf{rc} has finished, \textsf{init} launches virtual consoles
as usual, and each console runs a tty with \textsf{login} (through
\textsf{agetty}).

\lstset{language=sh, basicstyle=\footnotesize}
\begin{lstlisting}[frame=single]
c1:12345:respawn:/sbin/agetty 38400 tty1 linux
\end{lstlisting}

Note that when an \textsf{agetty} dies, \textsf{init} respawns it.

\vspace{1em}
Gentoo's \textsf{rc} program is currently being rewritten in C
(\textsf{baselayout} package version 2, see
\url{http://roy.marples.name/node/293}). The goal is to improve
performance and not to need bash any more.

\subsection{Configuration Management}
Runlevels are no longer named \textsf{rc[0-6S]}. Instead, they have an
alphanumerical name. Each runlevel has a corresponding directory in
\textsf{/etc/runlevels}, which contains symbolic links to scripts in
\textsf{/etc/init.d}.

The \textsf{rc-update} command is used to create, delete and modify
runlevels. Each time it is used, a service tree is recomputed. This
tree contains the list of services to start and their order, once all
dependencies have been computed and resolved.

The \textsf{rc-status} command shows the running status for each
service in a runlevel.

The \textsf{rc} command switches to a runlevel passed in argument.
Switching to a runlevel means comparing the list of running services
with the runlevel's service tree. If a service from
the service tree is not running, it is started. If a service is
running, but is neither in the service tree nor a dependency of a
service in the service tree, it is stopped. This does not apply when
leaving runlevels sysinit and boot.

\subsection{Life Cycle Management}
Each service has an init script particular to the Gentoo distribution.
Such a script at least understands \textsf{start}, \textsf{stop},
\textsf{restart}, \textsf{pause}, \textsf{zap}, \textsf{status},
\textsf{ineed}, \textsf{iuse}, \textsf{needsme}, \textsf{usesme} and
\textsf{broken} parameters. Life-cycle related parameters are:
\textsf{start} and \textsf{stop}, \textsf{restart} (with possibly extra
things to do between the stop and the re-start), \textsf{zap} (to
force the status to ``stopped'' in case
it is wrongly reported as ``stopped'')
and \textsf{status} (to query the running status).

Once a script is successfully \textsf{start}ed, its status is set to
``started''. This status will be set to
``stopped'' if the script is
\textsf{stop}ped or \textsf{zap}ped, and will be re-evaluated if the
script is \textsf{status}'ed. Re-evaluation is done
by checking the pid file in \textsf{/var/run} and the daemon file in
\textsf{/var/lib/init.d}. This means that if a
service's process is manually killed, the
service's status stays ``running'' until a \textsf{stop}, \textsf{zap} or
\textsf{status} command is issued.

\subsection{Dependency Resolution}
Gentoo init scripts respond to the following parameters: \textsf{ineed},
\textsf{iuse}, \textsf{needsme}, \textsf{usesme} and \textsf{broken}.
The first two list the hard and soft dependencies of the scripts, the
next two list the scripts that are hard- and soft-dependent on this
script, and the last one lists broken dependencies (unresolved,
``need''-type).

Inside the script, the following metadata tags can be used:
\textsf{need}, \textsf{use}, \textsf{before}, \textsf{after},
\textsf{provide}. \textsf{need} are hard dependencies: the current
script cannot run without them. \textsf{use} are soft dependencies: the
current script uses them if they are present in the current runlevel or
in boot, but can still run without them. \textsf{before} and
\textsf{after} specify a startup order, regardless of dependencies.
Finally, provide indicates virtual service names that the current
scripts fulfills (like in Gooch's need-based init).
All these tags are used when a script is manually invoked, and when a
runlevel is modified.

Network dependencies are a special case. A service will need a network
interface to be up (any interface other than net.lo). This dependency
is written: \textsf{need net}.

\subsection{Service Management Model}
All services must have a Gentoo-specific wrapper script (an additional
effort for the packager). Configurations and services can be managed
through good CLI tools.

For each service, settings (preferencess, configuration) are separated
from the wrapper script. Scripts are in
\textsf{/etc/init.d/{\textless}script{\textgreater}}, while settings are
in \textsf{/etc/conf.d/{\textless}script{\textgreater}}.

The main problem is that \textsf{rc-status} does not show the exact
state of services (cf. killed processes not seamlessly taken into
account).

\subsection{Service Management Tools}
\textsf{rc-status}, \textsf{rc-config}, \textsf{rc-update}: see above.
\\
\textsf{/sbin/rc}: switch runlevel (sys-apps/baselayout package).
\\
\textsf{depscan.sh}: rebuild the dependency tree.

\clearpage\section{InitNG}
\label{sec:initng}
\subsection{Home Page}
\url{http://www.initng.org/wiki}

\subsection{About}
InitNG's goal is to start services in parallel, for a
reduced startup time.

As a side note, the tool's syntax is reversed compared to System V:
instead of calling \textsf{{\textless}script{\textgreater} start}, InitNG
is invoked via \textsf{ngc start {\textless}script{\textgreater}}. It
makes more sense to call a single tool to manage life cycles than to call
multiple scripts that duplicate similar behaviour.

The configuration files support
lots of fancy features, such as respawn mode, suid, niceness, output
redirect, arguments passing, environment variables passing, etc.

There is not much information on the wiki, except that it is supposed to
be faster than SysVinit.

\subsection{Configuration Management}
\textsf{ng-update} manages configurations similarly to Gentoo's
\textsf{rc-update}.
\\
The wiki does not provide more information.

\subsection{Life Cycle Management}
Respawn mode for daemons. Otherwise, similar to SysVinit.

\subsection{Dependency Resolution}
\textsf{need} and \textsf{use} dependencies, similar to Gentoo init.
\\
Network access is a special dependency, as in Gentoo init.

\subsection{Service Management Model}
Every command goes through an InitNG program, instead of directly
invoking \textsf{rc} shell scripts.

\subsection{Service Management Tools}
The \textsf{ngc} tool is used to start / stop / restart services.
\\
The \textsf{ng-update} tool manages runlevels.

\clearpage\section{runit }
\label{sec:runit}
\subsection{Home Page}
\url{http://smarden.org/runit/}

\subsection{About}
From the home page: \textsf{runit} is a cross-platform Unix
init scheme with service supervision. Running and
shutting down is done in three stages:

\begin{enumerate}
\item \textbf{Stage 1:}\newline
\textsf{runit} starts \textsf{/etc/runit/1} and waits for it to
terminate. The system's one time initialization tasks
are done here. \textsf{/etc/runit/1} has full
control over \textsf{/dev/console} to be able to start an emergency
shell in case the one time initialization tasks fail.
\item \textbf{Stage 2:}\newline
\textsf{runit} starts \textsf{/etc/runit/2} which should not
return until the system is going to halt or reboot; if it crashes, it
will be restarted. Normally, \textsf{/etc/runit/2} runs
\textsf{runsvdir}.
\item \textbf{Stage 3:}\newline
If \textsf{runit} is told to halt or reboot the system, or
Stage 2 returns without errors, it terminates Stage 2 if it is running,
and runs \textsf{/etc/runit/3}. The systems tasks to shutdown and halt
or reboot are done here.
\end{enumerate}
\textsf{runit} is ``optimized for reliability
and small size''. If compiled and statically linked
with dietlibc, it produces an $\approx$ 8.5k executable.

\subsection{Configuration Management}
Runlevels are handled through the \textsf{runsvdir} and
\textsf{runsvchdir} programs.
\\
A runlevel is a directory in \textsf{/etc/runit/runsvdir/}. Symlinks
indicate which are the current and previous runlevel. A runlevel is
launched by \ \textsf{runsvdir}. Runlevel switching is performed by
\textsf{runsvchdir}, which switches directories for \textsf{runsvdir}.

\subsection{Life Cycle Management}
Each of a runlevel's subdirectories is a service. \ For
each service, \textsf{runsvdir} launches a \textsf{runsv}, which starts
and monitors this specific service. A service contains a ./run and an
optional ./finish script. Its state is maintained in a file by
\textsf{runsv}. A named pipe allows to pass commands to \textsf{runsv},
e.g. to start, stop, run once or in respawn mode, pause (SIGSTOP), etc.
\\
The \textsf{sv} program sends commands to one or all \textsf{runsv}s.

\subsection{Dependency Resolution}
Expression of dependencies is poor. If service A depends on service B,
A's run script must explicitly invoke sv to ensure
that B is running, e.g.:

\lstset{language=sh, basicstyle=\footnotesize}
\begin{lstlisting}[frame=single]
#!/bin/sh
sv start B || exit 1
exec A
\end{lstlisting}

There are no optional dependencies, no virtual service names, and no tools
to explore dependencies.

\subsection{Service Management Model}
One management process is launched for each service. A single runlevel
process is launched to manage all these management processes.
\\
Commands are passed through named pipes.

\subsection{Service Management Tools}
\textsf{sv}: pass life cycle-related commands from the CLI.
\\
\textsf{runsvchdir}: switch runlevel.
\\
\textsf{chpst} (change process state): set limits for a service or for a
user (depends on the limit, not really explicit). e.g. \# open files
per process, \# processes per user, memory per process.

\clearpage\section{Apple launchd }
\label{sec:launchd}
\subsection{Home Page}
\url{http://developer.apple.com/macosx/launchd.html}

\subsection{About}
Starting from Mac OS X 10.4, \textsf{launchd} aims to replace
\textsf{init}, the old \textsf{mach\_init}, \textsf{cron} and
\textsf{xinetd}. It starts daemons, shuts them down and restarts them
when a request comes (\textsf{inetd} behaviour), runs programs at a
specified time or later (\textsf{anacron}-like behaviour).

Each job (service) has an XML property file, replacing the different
syntaxes from crontab, inittab etc. This includes program arguments,
I/O redirect, setuid, chroot, etc.

Dependencies are automatically discovered (?). Services are run in
parallel.
\\
\textsf{launchd} can be controlled from a single utility,
\textsf{launchctl}.
\\
The source code has been relicensed under the ASL in august 2006.

\subsection{Configuration Management}
Runlevels are not supported on Mac OS (no order maintained, no selective
startup of services).
Directories of jobs can be used to launch a set of jobs together.

\subsection{Life Cycle Management}
Jobs can be started once, started in respawn mode, started in
launch-on-demand mode, started based on a schedule, stopped. How a
job is started is decided by its property file.

As usual, respawn mode requires that the service does not daemonize
itself (fork+exec, fork+exit).

\subsection{Dependency Resolution}
Dependencies are a bit unusual and lack documentation. The Uses,
Requires and Provides keywords have been dropped. Instead, services
watch for changes on a file or watch for IPCs. The advantage is that we
ensure dependencies are not only launched, but are also in a usable
state, already initialized and ready to communicate. The drawback is
that it is totally intrusive and quite obscure.

Special dependencies can be specified, also in the source code (not
as metadata), but using specific APIs: disk volume mounted, network
available, kernel module present (else it is launched), user Y is
logged in.

\subsection{Service Management Model}
Everything life cycle-related goes through \textsf{launchd}, instead
of using various tools (\textsf{cron}, \textsf{inet}). Every command is
issued using \textsf{launchctl}, which also has a prompt mode.
\\
Instances of \textsf{launchd} can be run for a specific user. This
allows non-root users to schedule jobs.
\\
No configuration management is provided, nor virtual service names.

\subsection{Service Management Tools}
\textsf{launchctl}: takes commands and passes them to \textsf{launchd}
through a mutex.

\clearpage\section{Ubuntu upstart }
\label{sec:upstart}
\subsection{Home Page}
\url{http://upstart.ubuntu.com/}

\subsection{About}
\textsf{upstart} aims to replace \textsf{init}, \textsf{cron},
\textsf{anacron}, \textsf{at} and \textsf{inet}. It also aims to
interact with \textsf{udev}, \textsf{acpi} and
\textsf{module-init-tools} (modprobe, etc). The basic idea is that
every action is triggered by an event.

Tasks (or jobs, or services) begin in the ``waiting'' state.
Upon reception of a task-dependent set of
events, the task is run and goes in the ``running''
state. Upon reception of another task-dependent
set of events, it goes to the ``dead''
state (which should be ``stopping''),
cleans up, then returns to the ``waiting''
state.

An event can be triggered manually, by ACPI, by \textsf{udev}, by
\textsf{upstart}, etc. Dependencies are also expressed as events:
\textsf{upstart} sends an event when a task goes ``running''; its
dependencies simply wait for this event.
Everything task-related is configured in \textsf{/etc/event.d}.

Security around events and event senders is unclear.

The project has been started before \textsf{launchd} went open source, and
shares some similarities (mainly \textsf{inet} and \textsf{cron}
capabilities).

Both code and documentation are quickly progressing, so some limitations
presented here may already have been fixed.

\subsection{Configuration Management}
A notion of profile is intended, as a replacement for runlevels. It is
still unspecified and unimplemented. Runlevels are still present as
dependencies: ``start on runlevel-2''.

\textsf{upstart} launches all tasks present in \textsf{/etc/event.d};
the order is defined by dependencies (expressed as events).

\subsection{Life Cycle Management}
Tasks can be waiting, running or dying. A set of events triggers
state switching. Actions can be performed at each state change
(start script, stop script, anything).

\subsection{Dependency Resolution}
Dependencies are expressed as e.g. ``start on event1''.  Event1 can be
that a service has begun to start (``on starting s1''), has finished to
stop (``on stopped s2''), etc. This unusual approach gives a lot of
flexibility: it allows virtual service names, optional dependencies, and a
lot more.

\subsection{Service Management Model}
The event system is very powerful and flexible. However, it might be a
bit complex to get the whole picture of what happens on the system.

\subsection{Service Management Tools}
\textsf{start}, \textsf{stop} and \textsf{status} commands are provided
to manage tasks.
\\
\textsf{initctl} is used to throw events from the command line (using a
mutex).

\clearpage\section{eINIT }
\label{sec:einit}
\subsection{Home Page}
\url{http://einit.sourceforge.net/}

\subsection{About}
eINIT aims to be a dependency-based init, written entirely in C,
suitable for embedded systems. It wants to get rid of rc scripts, to
have flexible dependencies, to be modular (e.g. with audio output for
the blind), and to support respawn mode.

\subsection{Configuration Management}

eINIT supports runlevels called modes. A mode is described by an XML
file. It can depend on other modes (like Gentoo's boot runlevel),
i.e. it comes in addition to the mode it depends on.
A mode says which services must be enabled (started), disabled
(stopped), and reset (restarted). A mode also defines orders of
precedence: ``try to start service A, and start
service B if it fails''. Finally, there is the notion of
service groups, e.g. all network interfaces, or all graphical devices.
Dependencies can be set on one, all or part of the service group.

Variables can be set globally and per-mode. (It is unclear what the use
case is)

\subsection{Life Cycle Management}
Services are either started once, started in respawn mode, or stopped.

Dependencies are expressed inside runlevels (cf. above), not in service
wrapper scripts. Services are described in the mode's XML file, with
attributes such as ``enable'' and ``disable'' that indicate what to run
when the service is started and stopped.

\subsection{Dependency Resolution}

Other attributes are ``provides'' and ``requires'' for dependencies. This
allows for virtual service names.  This also means that the XML
description must be duplicated between modes, unless we create a mode
dependency.

\subsection{Service Management Model}
Most management activities can be seen as switching runlevels.

eINIT supports enabling/disabling eINIT modules (e.g. TTYs, shell
commands, daemons), but not services themselves. This is a much coarser
granularity.

\subsection{Service Management Tools}
\textsf{einit-control} passes Commands through a Unix socket.

\clearpage\section{Pardus}
\label{sec:pardus}
\subsection{Home Page}
\url{http://www.pardus.org.tr/eng/projeler/comar/SpeedingUpLinuxWithPardus.html}

\subsection{About}
Pardus is a GNU/Linux distribution from Turkey. Two subprojects are within our
scope: Mudur and \c{C}omar.

Mudur is the init subsystem, which replaces the \textsf{rc} part of
\textsf{init}. It is called by \textsf{init} via the \textsf{/etc/inittab}
file. Mudur performs runlevel switching tasks, according to the usual
System V runlevels (boot, single, reboot, etc). Mudur is written in
Python.

\c{C}omar is the "configuration management system" which manages settings
and preferencesi, and handles the way to start applications. It needs
modified rc scripts, as do most of its competitors. According to the
wiki, it "handles parallel execution, access control, profile management
and remote management". \c{C}omar is written in C, and rc scripts are
written in Python.

The main objective is to gain speed at startup, while being clean and
hiding complexity to the user.

\subsection{Configuration Management}
Runlevels have the same meaning as in System V. For the sysinit runlevel,
Mudur mounts pseudo-filesystems, starts \textsf{udev}, loads kernel
modules, etc. For other runlevels, Mudur calls \c{C}omar, which in turn
handles rc scripts.

\subsection{Life Cycle Management}
rc scripts accept parameters such as start, stop, status, configure. The
start and stop directives make use of Debian's \textsf{start-stop-daemon}.

\subsection{Dependency Resolution}
An rc script calls the start methods of the other scripts it depends on.
\\
There are no virtual service names.

\subsection{Service Management Model}
Command-line tools allow to call methods on rc scripts via \c{C}omar.

\subsection{Service Management Tools}
Not presented.

\clearpage\section{LSB init Specifications }
\label{sec:lsb}
\subsection{Home Page}
\url{http://www.freestandards.org/spec/refspecs/LSB\_1.3.0/gLSB/gLSB/sysinit.html}

\subsection{About}
LSB, the Linux Standards Base, defines what init scripts should conform
to.
\\
This is an opinion, but the LSB seems more keen on standardizing what
RedHat does than on being innovative.

\subsection{Configuration Management}
LSB specifies the following runlevels:

0: halt

1: single user mode

2: multiuser with no network services exported

3: normal/full multiuser

4: reserved for local use, default is normal/full multiuser

5: multiuser with xdm or equivalent

6: reboot
\\
This forbids ``profiles'' and useful runlevel switching as seen in
previous init schemes (e.g. Gentoo).

\subsection{Life Cycle Management}
Init files provided by LSB applications shall accept one argument,
which states the action to perform:

\textbf{start}: start the service

\textbf{stop}: stop the service

\textbf{restart}: stop and restart the service if the service is already
running, otherwise start the service

\textbf{reload}: cause the configuration of the service to be reloaded
without actually stopping and restarting the service [optional]

\textbf{force-reload}: cause the configuration to be reloaded if the
service supports this, otherwise restart the service

\textbf{status}: print the current status of the service
\\
There is no respawn mode.

\subsection{Dependency Resolution}
LSB-compliant applications init scripts (but not system init scripts)
must adopt this syntax for dependency declaration:

\lstset{language=sh, basicstyle=\footnotesize}
\begin{lstlisting}[frame=single]
# Provides: boot_facility_1 [ boot_facility_2 ...]
# Required-Start: boot_facility_1 [boot_facility_2 ...]
# Required-Stop: boot_facility_1 [boot_facility_2 ...]
# Default-Start: run_level_1 [ run_level_2 ...]
# Default-Stop: run_level_1 [ run_level_2 ...]
# Short-Description: short_description
# Description: multiline_description
\end{lstlisting}

Required-stop are applications that must still run before this one
stops.

The ``Default-Start'' and ``Default-Stop'' headers define which runlevels
should by default run the script with a start or stop argument,
respectively, to start or stop the services controlled by the init script.

``Provides'' allows to have virtual service names.

\subsection{Service Management Model}
Out of scope.

\subsection{Service Management Tools}
Out of scope.

\clearpage\section{Daemontools: svscan }
\label{sec:daemontools}
\subsection{Home Page}
\url{http://code.dogmap.org/svscan-1/}
\\
\url{http://cr.yp.to/daemontools.html}

\subsection{About}
From the daemontools page:
\textsf{daemontools} is a collection of tools for managing UNIX
services.
The \textsf{svscan} program can be used as an init replacement.

\subsection{Configuration Management}
Configurations / runlevels are supported as directories of \textsf{run}
scripts.

\textsf{svscan} starts and monitors a collection of services.
\textsf{svscan} starts one \textsf{supervise} process for each
subdirectory of the current directory. Every five seconds,
\textsf{svscan} checks for subdirectories again. If it sees a new
subdirectory, it starts a new \textsf{supervise} process.

Runlevel switching is not really specified.

\subsection{Life Cycle Management}
\textsf{supervise} monitors a service. It starts the service and
restarts it if it dies. To set up a new service, all \textsf{supervise}
needs is a directory with a \textsf{run} script that runs the service.

\textsf{supervise} maintains status information in a binary format inside
the directory
\textsf{\textit{{\textless}service{\textgreater}}/supervise}, which must
be writable by \textsf{supervise}. The status information can be read
using \textsf{svstat}.

There is no stop script.

\subsection{Dependency Resolution}
No dependencies.

\subsection{Service Management Model}
Put run scripts in a directory; daemontools will run the service and
monitor it.
\\
It is simpler than SysVinit, but lacks features (stop command, runlevel
switching).

\subsection{Service Management Tools}
\textsf{svc} is used to send signals to a service.
\\
\textsf{svstat} prints the status of services monitored by
\textsf{supervise}.
\\
\textsf{multilog} saves error messages to one or more logs. It
optionally timestamps each line and, for each log, includes or excludes
lines matching specified patterns.

\clearpage\section{minit }
\label{sec:minit}
\subsection{Home Page}
\url{http://www.fefe.de/minit/}

\subsection{About}
Minit aims to be a very small init.
It can start a service, restart it, start it when another is finished,
and take dependencies into account.

\subsection{Configuration Management}
None.

\subsection{Life Cycle Management}
Reads \textsf{/var/run/{\textless}service{\textgreater}.pid} and checks
if the process has died.

\subsection{Dependency Resolution}
Yes, but lacks documentation.

\subsection{Service Management Model}
None.

\subsection{Service Management Tools}
\textsf{msvc}: similar to svc from daemontools.
\\
\textsf{pidfilehack}: reads
\textsf{/var/run/{\textless}service{\textgreater}.pid}.

\clearpage\section{jinit }
\label{sec:jinit}
\subsection{Home Page}
\url{http://john.fremlin.de/programs/linux/jinit/}

\subsection{About}
jinit is based on an extended version of Richard
Gooch's need(8) scheme for init. Currently there is
no provide(8) functionality.

\subsection{Configuration Management}
None? (or at least documentation is lacking).

\subsection{Life Cycle Management}
Supports respawn mode. Apart from that, same as SysVinit.

If it receives a fatal error signal like SIGSEGV it will fork a child to
dump core, and exec itself (with the ``panic'' option so no bootscripts
are run).  That means it should never bring the system down unnecessarily.

\subsection{Dependency Resolution}
Same as the need(8) scheme.

\subsection{Service Management Model}
Same as SysVinit.

\subsection{Service Management Tools}
None. Communicates over SysV message queues.

\clearpage\section{cinit }
\label{sec:cinit}
\subsection{Home Page}
\url{http://linux.schottelius.org/cinit/}

\subsection{About}
\textsf{cinit} claims to be a fast, small and simple init with
support for profiles. It supports parallel startup,
mandatory and optional dependencies, and does seem simple to use.

Size is $\approx$ 50KiB when statically linked against uclibc.

\subsection{Configuration Management}
``Profiles'' are directories of services.
They are put in \textsf{/etc/cinit}.
\\
Choosing the profile at startup is supported, but profile switching is
not documented.

\subsection{Life Cycle Management}
A service is a subfolder in a configuration (or a link to another
folder). Everything is done through files or symlinks: \textsf{./on}
and \textsf{./off} indicate the program to run when the service goes
on/off. \textsf{./on.params} and \textsf{./on.env} contain parameters
and environment variables to pass to the program.

\subsection{Dependency Resolution}
Dependencies are listed in ./wants (optional) and \textsf{./needs}
(mandatory).
\\
Virtual service names are not supported. The author suggests using
symlinks, e.g. \textsf{cron~-{\textgreater}~dcron}.

\subsection{Service Management Model}
None.

\subsection{Service Management Tools}
A bunch of executables:

\lstset{language=sh, basicstyle=\footnotesize}
\begin{lstlisting}[frame=single]
cinit.add.dependency - add a dependency to a service
cinit.add.getty      - add a new getty
cinit.create.empty.service - create an empty service
cinit.reboot         - reboot in /bin/sh
cinit.remove.getty   - remove a getty service
cinit.respawn.off    - switch respawing off
cinit.respawn.on     - switch respawing on
cinit.shutdown       - shutdown in /bin/sh
\end{lstlisting}

\clearpage\section{monit }
\label{sec:monit}
\subsection{Home Page}
\url{http://www.tildeslash.com/monit/index.php}

\subsection{About}
\textsf{monit} manages and monitors processes, files, directories and
devices. It does not replace init: instead, it acts as a service
manager, and can reuse rc scripts.

\textsf{monit} can start a process which is not running, restart a process
if it does not respond, and stop a process if it uses too much resources.
It can monitor files, directories and devices for changes on timestamp,
checksum or size. All checks are performed at a polling interval time.

Management activities are exported through HTTP(S).
Monit is configured via the \textsf{/etc/monitrc} file.

\subsection{Configuration Management}
Only supports groups of services which can be started / stopped together.

\subsection{Life Cycle Management}
A service can be (re)started and stopped. Its state is checked using pid
files. This means that wrapper scripts may have to be written.

\subsection{Dependency Resolution}
The syntax is simplistic: \textsf{depends on serviceX} (in file
\textsf{/etc/monitrc}).
\\
There are no virtual service names.

\subsection{Service Management Model}
Periodically check a resource (process, file\ldots). Upon changes, actions
are triggered. For example:

\lstset{language=sh, basicstyle=\footnotesize}
\begin{lstlisting}[frame=single]
  check process apache with pidfile /var/run/httpd.pid
        start program = ``/etc/init.d/httpd start''
        stop program  = ``/etc/init.d/httpd stop''
        if failed host www.tildeslash.com port 80 then restart
        depends on apache_bin
\end{lstlisting}

\subsection{Service Management Tools}
HTTP interface.

\clearpage\section{depinit}
\label{sec:depnit}
\subsection{Home Page}
\url{http://www.nezumi.plus.com/depinit/}

\subsection{About}
From the home page: "Depinit is an alternative init program that can
handle parallel execution, dependencies, true roll-back, pipelines,
improved signaling and unmounting filesystems on shutdown. It incorporates
ideas from sysvinit, simpleinit, daemontools and make. At present, it is a
bit experimental, and requires good knowledge of the initialisation
process to set up."

While depinit boasts more features than sysvinit, it is quite complex to
setup and to use. It also has trouble working with the \textsf{login}
program.

\subsection{Configuration Management}
A runlevel is a directory in \textsf{/etc/depinit}. How it operates and
how it can be switched is unclear.

\subsection{Life Cycle Management}
Each service has its own directory in \textsf{/etc/depinit}, which will
contain shell scripts named start, stop, source, filter, dest and/or
daemon. Their purpose is either obvious (start and stop describe shell
commands that will be called to start and stop the service), or
unexplained.

\subsection{Dependency Resolution}
Dependencies are listed in a \textsf{depend} file within each service's
directory. Each line in the file corresponds to another directory in
\textsf{/etc/depinit}.

\subsection{Service Management Model}
All interactions with depinit are performed through the \textsf{depctl}
command-line tool. Daemons are started in respawn mode.

\subsection{Service Management Tools}
\textsf{depinit}: can either replace init (and rc), or be started with a
PID {\textgreater} 1 and manage a subset of services.
\\
\textsf{depctl}: start/stop a service, shutdown/reboot the system, send a
signal to a daemon.
\\
These tools communicate via signals.

\clearpage\section{NetBSD rc.d }
\label{sec:netbsd}
\subsection{Home Page}
\url{http://ezine.daemonnews.org/200108/rcdsystem.html}
\\
\url{http://www.netbsd.org/guide/en/chap-rc.html}
\\
\url{http://cvsweb.netbsd.org/bsdweb.cgi/src/sbin/init/}

\subsection{About}
From the home page:
\\
As of NetBSD 1.5, the startup of the system changed to using
rc-scripts for controlling services, similar to SysVinit, but without
runlevels.

\subsection{Configuration Management}
None.

\subsection{Life Cycle Management}
Service wrapper scripts understand \textsf{start}, \textsf{stop},
\textsf{restart} and \textsf{kill}.
\\
They are as limited as SysVinit's. There is no service monitoring
facility.

\subsection{Dependency Resolution}
Dependencies are expressed in rc scripts as four keywords. REQUIRE is a
mandatory dependency; PROVIDE allows virtual service names;
BEFORE's explanation fell through a spatiotemporal
void; KEYWORD is a tag used to include or exclude the script from
ordering, e.g. with values 'nostart' or 'shutdown'.

Ordering is done via the \textsf{rcorder} command.

\subsection{Service Management Model}
None. Start scripts, hope they run, optionally check a pid file.

\subsection{Service Management Tools}
None.

\clearpage\section{Solaris SMF }
\label{sec:smf}
\subsection{Home Page}
\url{http://www.sun.com/bigadmin/content/selfheal/smf-quickstart.html}

\subsection{About}
SMF (Service Management Facility) is a part of SUN Solaris 10
(historically a System V variant). It targets big production servers.
From their page:

SMF has added the notion of the relationship between a service, its
processes, and another service that is responsible for restarting the
service, to the Solaris kernel. SMF understands whether a service
process failed as the result of an administrator error, failure of a
dependent service, software bug, or underlying hardware failure. SMF
informs the appropriate restarter, which decides either to disable the
service by placing it in \textup{maintenance mode} because it appears
to be defective, or to restart it \textup{automatically}.

SMF still supports \textsf{inittab}, \textsf{inetd} and rc scripts, but
aims at replacing them.

SMF services are started in parallel, in a dependency-based order.

Users can be given the rights to manage services and their state,
without being root. Still, there are no per-user services.

Alas, the source code is closed.

\subsection{Configuration Management}
Runlevels are replaced by milestones. A specific set of services must be
enabled (started) before a milestone is reached (incremental
runlevels). Milestones are: single-user, multi-user,
multi-user-server, all, none.

The notion of profile also exist: it is an XML file which lists SMF
services, and whether each should be enabled or disabled.

\subsection{Life Cycle Management}
SMF-enabled services can be permanently enabled / disabled (persists
on reboot). When enabled, they can be online (started), offline
(stopped), uninitialized (config has not been read yet), degraded (runs
at limited capacity), and maintenance (stopped, needs an action from
the administrator).

Possible commands are: permanent or temporary enable / disable, start,
stop, restart, refresh. Refresh means reloading a
service's config file, which is usually triggered by
SIGHUP on daemons (SIGHUP should kill non-daemon processes).

\subsection{Dependency Resolution}
A dependency is formed as an URI. It can be set on services, devices,
network configuration, kernel configuration, milestone (runlevel), and
be optional or mandatory. Shortened example from SUN's
site:

\lstset{language=sh, basicstyle=\footnotesize}
\begin{lstlisting}[frame=single]
% svcs -l network/smtp:sendmail
dependency  require_all/refresh file://localhost/etc/mail/sendmail.cf (-)
dependency  optional_all/none svc:/system/system-log (online)
dependency  require_all/refresh svc:/system/identity:domain (online)
dependency  require_all/refresh svc:/milestone/name-services (online)
dependency  require_all/none svc:/network/service (online)
dependency  require_all/none svc:/system/filesystem/local (online)
\end{lstlisting}

\subsection{Service Management Model}
SMF manages services, rc scripts, inet daemons, configurations in an
(almost) uniform way. The goal is to obtain the best uptime possible,
with as little human intervention as possible. The other big goal is to
provide detailed information on failures to the administrator, in the
case he must intervene.

There are lots and lots of features, with probably lots of complicated
code: even the kernel is modified. This solution is ideal for servers,
but not suitable for small systems.

\subsection{Service Management Tools}
Amongst others:

\textsf{scvs} queries the system state, including old rc scripts.

\textsf{svcadm} passes commands to SMF services and their restarters.

\textsf{svccfg} switches profiles.

\clearpage\section{Other Contenders}
\label{sec:others}
The following projects are considered less relevant. Most are dead
and/or lack documentation.

\subsection{DMD}
\label{sec:dmd}

Daemons-Managing Daemon is a stalled project from the Free Software
Foundation. Documentation is non-existent.

\url{http://directory.fsf.org/GNU/DMD.html}

\subsection{twsinit}
\label{sec:twsinit}

\textsf{twsinit} is a minimal init replacement (about 8 KB) in x86 assembler.
Documentation is non-existent.

\url{http://www.energymech.net/users/proton/}

\subsection{Busybox}
\label{sec:busybox}

BusyBox combines tiny versions of many common UNIX utilities (GNU
fileutils, shellutils, textutils) into a single small executable. It
contains a small init system, which is a stripped down SysVinit scheme.
It can read an \textsf{inittab} file, but does not support runlevels.

The project is alive.

\url{http://www.busybox.net/}

\subsection{Fedora New Init System}
\label{sec:fedora}

Fedora had planned to implement its own init replacement. Its goals were
a bit confused and mixed: better startup speed, LSB compliance,
respawn, logging, tools. The project is stalled / dead.

\url{http://fedoraproject.org/wiki/FCNewInit}

\subsection{serel}
\label{sec:serel}

Serel aims to speed up the boot sequence using parallel startup. The
project is stalled.

\url{http://www.fastboot.org/}

%%%
%%% Commented: wajig relates to the package manager, not to init
%%%
%\subsection{Debian wajig}
%\label{sec:wajig}
%
%Written in Python, this tool centralizes interactions with
%\textsf{apt-get}, \textsf{dpkg}, \textsf{apt-cache}, \textsf{wget}, etc.
%It gives access to package deployment (install, update, etc), package
%creation, package dependency resolution, package information (readme,
%bugs, or anything written inside package metadata / repository metadata),
%application life cycle (start / stop / restart rc scripts), and Debian
%preferences such as setting default applications.
%
%\url{http://www.togaware.com/wajig/}

\clearpage
\begin{landscape}
{\tabfont
\begin{longtable}[c]{|p{1.8cm}|p{2.1cm}|p{2.5cm}|p{2.9cm}|p{2.5cm}|p{2.7cm}|p{1.2cm}|}
\hline
&
{\centering
\textbf {Focus}
\par}
&
{\centering
\textbf {Configuration Management}
\par}
&
{\centering
\textbf {Life Cycle Instrumentation}
\par}
&
{\centering
\textbf {Life Cycle Monitoring}
\par}
&
{\centering
\textbf {Expression of Dependencies}
\par}
&
{\centering
\textbf {License}
\par}
\\\hline
\endhead
{\raggedleft
\textbf {BSD init}
\par}
&
{\centering
-
\par}
&
{\centering
No
\par}
&
{\centering
No
\par}
&
{\centering
No
\par}
&
{\centering
Static ordering
\par}
&
{\centering
BSD
\par}
\\\hline
{\raggedleft
\textbf {SysVinit}
\par}
&
{\centering
-
\par}
&
{\centering
Numbered runlevels
\par}
&
{\centering
start, stop
\par}
&
{\centering
Rc scripts (/var)
\par}
&
{\centering
Static ordering
\par}
&
{\centering
GPL
\par}
\\\hline
{\raggedleft
\textbf {simpleinit}
\par}
&
{\centering
Have dependencies
\par}
&
{\centering
No
\par}
&
{\centering
start, stop
\par}
&
{\centering
Rc scripts (/var)
\par}
&
{\centering
need
\par}
&
{\centering
GPL
\par}
\\\hline
{\raggedleft
\textbf {Gentoo rc}
\par}
&
{\centering
Better runlevels and dependencies
\par}
&
{\centering
Named runlevels
\par}
&
{\centering
start, stop, restart
\par}
&
{\centering
Rc scripts (/var)
\par}
&
{\centering
need, use, before, after, provide
\par}
&
{\centering
GPL
\par}
\\\hline
{\raggedleft
\textbf {InitNG}
\par}
&
{\centering
Speed
\par}
&
{\centering
Named runlevels
\par}
&
{\centering
start, stop, respawn
\par}
&
{\centering
Process (/var)
\par}
&
{\centering
need, use
\par}
&
{\centering
GPL
\par}
\\\hline
{\raggedleft
\textbf {runit}
\par}
&
{\centering
Life Cycle Management
\par}
&
{\centering
Named runlevels
\par}
&
{\centering
POSIX signals
\par}
&
{\centering
Father process\textit{es}
\par}
&
{\centering
manual scripting
\par}
&
{\centering
BSD
\par}
\\\hline
{\raggedleft
\textbf {launchd}
\par}
&
{\centering
Lots of features
\par}
&
{\centering
No
\par}

{\centering
(Groups of jobs)
\par}
&
{\centering
start, stop, inet, cron, respawn
\par}
&
{\centering
Father process
\par}
&
{\centering
In code: watch files \& IPCs
\par}
&
{\centering
ASL
\par}
\\\hline
{\raggedleft
\textbf {upstart}
\par}
&
{\centering
Event-based
\par}
&
{\centering
Profiles
\par}
&
{\centering
Events
\par}
&
{\centering
Father process
\par}
&
{\centering
stopping, stopped, starting, started, any event
\par}
&
{\centering
GPL
\par}
\\\hline
{\raggedleft
\textbf {eINIT}
\par}
&
{\centering
Runlevels
\par}
&
{\centering
Modes (services, dependencies, vars)
\par}
&
{\centering
start, stop, respawn
\par}
&
{\centering
Father process
\par}
&
{\centering
provides, requires
\par}
&
{\centering
BSD
\par}
\\\hline
{\raggedleft
\textbf {Pardus}
\par}
&
{\centering
Speed
\par}
&
{\centering
Numbered runlevels
\par}
&
{\centering
start, stop
\par}
&
{\centering
Rc scripts (/var)
\par}
&
{\centering
need
\par}
&
{\centering
GPL
\par}
\\\hline
{\raggedleft
\textbf {LSB}
\par}
&
{\centering
Standardize dependencies
\par}
&
{\centering
Numbered runlevels
\par}
&
{\centering
start, stop, restart
\par}
&
{\centering
Rc scripts
\par}
&
{\centering
provides, required-to-start, required-to-stop
\par}
&
{\centering
-
\par}
\\\hline
{\raggedleft
\textbf {svscan}
\par}
&
{\centering
Service Management
\par}
&
{\centering
Directories of run scripts
\par}
&
{\centering
start, respawn
\par}
&
{\centering
Father process\textit{es}
\par}
&
{\centering
No
\par}
&
{\centering
freedist
\par}
\\\hline
{\raggedleft
\textbf {minit}
\par}
&
{\centering
Size
\par}
&
{\centering
No
\par}
&
{\centering
\ respawn
\par}
&
{\centering
/var
\par}
&
{\centering
?
\par}
&
{\centering
GPL
\par}
\\\hline
{\raggedleft
\textbf {jinit}
\par}
&
{\centering
need(8) deps
\par}
&
{\centering
No
\par}
&
{\centering
\ respawn
\par}
&
{\centering
/var ?
\par}
&
{\centering
need
\par}
&
{\centering
GPL
\par}
\\\hline
{\raggedleft
\textbf {cinit}
\par}
&
{\centering
Size, runlevels
\par}
&
{\centering
Unswitchable profiles
\par}
&
{\centering
respawn
\par}
&
{\centering
Scripts (/var)
\par}
&
{\centering
wants, needs
\par}
&
{\centering
GPL
\par}
\\\hline
{\raggedleft
\textbf {monit}
\par}
&
{\centering
Service Management
\par}
&
{\centering
No
\par}

{\centering
(Groups of jobs)
\par}
&
{\centering
start, stop, restart, quotas
\par}
&
{\centering
Scripts (/var)
\par}
&
{\centering
depends
\par}
&
{\centering
GPL
\par}
\\\hline
%{\raggedleft
%\textbf {depinit}
%\par}
%&
%{\centering
%Performance tweaks
%\par}
%&
%{\centering
%Directories of scripts
%\par}
%&
%{\centering
%start, stop, respawn
%\par}
%&
%{\centering
%Father process?
%\par}
%&
%{\centering
%depend
%\par}
%&
%{\centering
%GPL
%\par}
%\\\hline
{\raggedleft
\textbf {NetBSD rc.d}
\par}
&
{\centering
Have rc scripts
\par}
&
{\centering
No
\par}
&
{\centering
start, stop, restart
\par}
&
{\centering
Rc scripts
\par}
&
{\centering
require, provide, before
\par}
&
{\centering
BSD
\par}
\\\hline
{\raggedleft
\textbf {SMF}
\par}
&
{\centering
Heavy Management
\par}
&
{\centering
Milestones and profiles
\par}
&
{\centering
start, stop, restart
\par}

{\centering
+ persist + self-healing
\par}
&
{\centering
Father Process / kernel
\par}
&
{\centering
require, optional
\par}

{\centering
on URIs (svc, file...)
\par}
&
{\centering
Closed source
\par}
\\\hline
\end{longtable}
}
\end{landscape}

%%%%%%%%%%%%%%%%%%%%%%%%%%%%%%%%%%%%%%%%%%%%%%%%%%%%%%%%%%%%%%%%
%%%%%%%%%%%%%%%%%%%%%%%%%%%%%%%%%%%%%%%%%%%%%%%%%%%%%%%%%%%%%%%%

\end{document}